%
% the following is to use blackboard bold fonts --
\let\useblackboard=\iftrue
%
% activate this if you don't have them.
%\let\useblackboard=\iffalse
%
% You might also need to remove this line.
\newfam\black
\input harvmac.tex
\noblackbox
\def\Title#1#2{\rightline{#1}
\ifx\answ\bigans\nopagenumbers\pageno0\vskip1in%
\baselineskip 15pt plus 1pt minus 1pt
\else%\special{papersize=11in,8.5in}%
\def\listrefs{\footatend\vskip
1in\immediate\closeout\rfile\writestoppt
\baselineskip=14pt\centerline{{\bf
References}}\bigskip{\frenchspacing%
\parindent=20pt\escapechar=` \input
refs.tmp\vfill\eject}\nonfrenchspacing}
\pageno1\vskip.8in\fi \centerline{\titlefont #2}\vskip .5in}
 
scaled\magstep3
 
scaled\magstep3
 
scaled\magstep3
 
scaled\magstep3
 
scaled\magstep3
\ifx\answ\bigans\def\tcbreak#1{}\else\def\tcbreak#1{\cr&{#1}}\fi
\useblackboard
\message{If you do not have msbm (blackboard bold) fonts,}
\message{change the option at the top of the tex file.}

\font\blackboard=msbm10 scaled \magstep1
\font\blackboards=msbm7
\font\blackboardss=msbm5
%\newfam\black
\textfont\black=\blackboard
\scriptfont\black=\blackboards
\scriptscriptfont\black=\blackboardss

\else

\fi
% *************************************
%\draftmode
%

%
\def\yboxit#1#2{\vbox{\hrule height #1 \hbox{\vrule width #1
\vbox{#2}\vrule width #1 }\hrule height #1 }}
\def\fillbox#1{\hbox to #1{\vbox to #1{\vfil}\hfil}}
\def\ybox{{\lower 1.3pt \yboxit{0.4pt}{\fillbox{8pt}}\hskip-0.2pt}}

\def\comments#1{}

\def\p{\partial}

\def\half{{1\over 2}}
\def\Tr{{{\rm Tr\  }}}

\def\vev#1{\langle{#1}\rangle}

\def\CF{{\cal F}}

\def\CM{{\cal M}}

\def\CK{{\cal K}}

\def\II{\relax{I\kern-.07em I}}

\def\inbar{\,\vrule height1.5ex width.4pt depth0pt}
\def\IZ{\relax\ifmmode\mathchoice
{\hbox{\cmss Z\kern-.4em Z}}{\hbox{\cmss Z\kern-.4em Z}}
{\lower.9pt\hbox{\cmsss Z\kern-.4em Z}}
{\lower1.2pt\hbox{\cmsss Z\kern-.4em Z}}\else{\cmss Z\kern-.4em
Z}\fi}
\def\IB{\relax{\rm I\kern-.18em B}}
\def\IC{{\relax\hbox{$\inbar\kern-.3em{\rm C}$}}}
\def\ID{\relax{\rm I\kern-.18em D}}
\def\IE{\relax{\rm I\kern-.18em E}}
\def\IF{\relax{\rm I\kern-.18em F}}
\def\IG{\relax\hbox{$\inbar\kern-.3em{\rm G}$}}
\def\IGa{\relax\hbox{${\rm I}\kern-.18em\Gamma$}}
\def\IH{\relax{\rm I\kern-.18em H}}
\def\IK{\relax{\rm I\kern-.18em K}}
\def\IP{\relax{\rm I\kern-.18em P}}
\def\pp{{\relax{=\kern-.42em |\kern+.2em}}}
%\def\IX{\relax{\rm X\kern-.01em X}}
%this doesn't work

\def\p{\partial}

\font\cmss=cmss10 \font\cmsss=cmss10 at 7pt
\def\IR{\relax{\rm I\kern-.18em R}}

\def\Tr{{\rm Tr\ }}

\def\Bone{{\bf 1}}

\def\str{{\rm ~STr~}}
\def\mX{{\bf X}}

%
%Journal macros
%

\def\NP{{\it Nucl. Phys.\ }}

\def\PL{{\it Phys. Lett.\ }}
\def\PR{{\it Phys. Rev.\ }}
\def\PRL{{\it Phys. Rev. Lett.\ }}

\def\JHEP{ {\it JHEP }}

\Title{ \vbox{\baselineskip12pt\hbox{hep-th/9802173}
\hbox{BROWN-HET-1113}
}}
{\vbox{
\centerline{Statistical Origin of Black Hole Entropy in}
\centerline{Matrix Theory}}}

\centerline{David A. Lowe}
\medskip
\centerline{Department of Physics}
\centerline{Brown University}
\centerline{Providence, RI 02912, USA}
\centerline{\tt lowe@het.brown.edu}
\bigskip

The statistical entropy of black holes in Matrix theory is considered.
Assuming Matrix theory is the discretized light-cone quantization of a
theory with eleven-dimensional Lorentz invariance, 
we map the counting problem
onto the original Gibbons-Hawking calculation of the thermodynamic
entropy.

\Date{February, 1998}

%References
\lref\antal{A. Jevicki and B. Sakita, ``The Quantum Collective Field
Method and its Application to the Planar Limit,'' \NP {\bf B165}
(1980) 511.}
\lref\deser{S. Weinberg, ``Photons and Gravitons in Perturbation
Theory: Derivation of Maxwell's and Einstein's Equations,'' \PR {\bf
B138} (1965) 988; S. Deser, ``Self-Interaction and Gauge Invariance,'' {\it
Gen. Rel. Grav.} {\bf 1} (1970) 9.}
\lref\gibhawk{G.W. Gibbons and S.W. Hawking, ``Action integrals and
partition functions in quantum gravity,'' \PR {\bf D15} (1977) 2752.}
\lref\frolov{V.P. Frolov and D.V. Fursaev, ``Plenty of Nothing: Black
Hole Entropy in Induced Gravity,'' hep-th/9705207.}
\lref\jacobson{T. Jacobson, ``Black Hole Entropy and Induced
Gravity,'' gr-qc/9404039.}
\lref\sakharov{A.D. Sakharov, ``Vacuum Quantum Fluctuations in Curved
Space and the Theory of Gravitation,'' {\it Sov. Phys. Doklady} {\bf
12} (1968) 1040; ``Spectral Density of Eigenvalues of the Wave
Equation and Vacuum Polarization,''
 {\it Theor. Math. Phys.} {\bf 23} (1976) 435.}
\lref\kabat{D. Kabat and W. Taylor, ``Linearized Supergravity from
Matrix Theory,'' hep-th/9712185.}
\lref\chepelev{I. Chepelev and A. Tseytlin, ``Long-distance
interactions of branes: correspondence between supergravity and
Yang-Mills descriptions,'' \NP {\bf B515} (1998) 73, hep-th/9709087; 
``Interactions of Type IIB
D-branes from D-instanton matrix model,'' \NP {\bf B511} (1998) 629, hep-th/9704127;
D. Kabat and W. Taylor, ``Spherical Membranes in Matrix Theory,''
hep-th/9711078.}
\lref\douglas{M.R. Douglas, ``Large N Gauge Theory - Expansions and
Transitions,'' hep-th/9409098.}
\lref\laflamme{R. Gregory and R. Laflamme, ``Black Strings and
p-Branes are Unstable,'' \PRL {\bf 70} (1993) 2837, hep-th/9301052.}
\lref\banks{T. Banks, ``Matrix Theory,'' hep-th/9710231.}
\lref\strovaf{A. Strominger and C. Vafa, ``Microscopic Origin of the
Bekenstein-Hawking Entropy,'' \PL {\bf B379} (1996) 99,
hep-th/9601029; C.G. Callan and J.M. Maldacena, ``D-brane Approach to
Black Hole Quantum Mechanics,'' \NP {\bf B472} (1996) 591, hep-th/9602043;
G.T. Horowitz and A. Strominger, ``Counting States of Near Extremal
Black Holes,'' \PRL {\bf 77} (1996) 2368, hep-th/9602051.}
\lref\matrixbh{T. Banks, W. Fischler, I.R. Klebanov and L. Susskind,
``Schwarzschild Black Holes in Matrix Theory I,II,''
\PRL {\bf 80} (1998) 226, hep-th/9709091; \JHEP {\bf 01} (1998) 008, 
hep-th/9711005; 
I.R. Klebanov and L. Susskind, 
``Schwarzschild Black Holes in Various Dimensions from Matrix
Theory,''
\PL {\bf B416} (1998) 62, hep-th/9709108; G.T Horowitz and E. Martinec, 
``Comments on Black Holes in Matrix Theory,''
\PR {\bf D57} (1998) 4935, hep-th/9710217; M. Li, 
``Matrix Schwarzschild Black Holes in Large N
limit,'' \JHEP {\bf 01} (1998) 009, hep-th/9710226;
N. Ohta and J-G Zhou, ``Euclidean Path Integral, D0-Branes and
Schwarzschild Black Holes in Matrix Theory,'' hep-th/9801023.}
\lref\ooguri{M.R. Douglas and H. Ooguri, ``Why Matrix Theory is
Hard,''
hep-th/9710178.}
\lref\seiberg{N. Seiberg, ``Why is the Matrix Model Correct?'' \PRL
{\bf 79} (1997) 3577, hep-th/9710009; A. Sen, ``D0 Branes on $T^n$ 
and Matrix Theory,'' hep-th/9709220.}
\lref\fiola{T.M. Fiola, J. Preskill, A. Strominger and S.P. Trivedi,
``Black hole thermodynamics and information loss in two dimensions,''
\PR {\bf D50} (1994) 3987, hep-th/9403137.}
\lref\fursaev{V.P. Frolov, D.V. Fursaev and A.I. Zelnikov, 
``Statistical Origin Of Black Hole Entropy In Induced Gravity,'' \NP
{\bf B486} (1997) 339, hep-th/9607104;
V.P. Frolov and D.V. Fursaev, ``Mechanism Of Generation
Of Black Hole Entropy In Sakharov's Induced Gravity,'' \PR {\bf D56}
(1997) 2212, hep-th/9703178.}
\lref\sethi{S. Paban, S. Sethi and M. Stern, ``Constraints from Extended Supersymmetry in
Quantum Mechanics,'' hep-th/9805018;
M. Dine, R. Echols and J. Gray, ``Renormalization of Higher Derivative 
Operators in the Matrix Model,'' hep-th/9805007.}

\newsec{Introduction}

Recently, Matrix theory was proposed as the first
nonperturbative formulation of string theory. 
It may be thought of
as a quantization of eleven-dimensional M-theory,
with a light-like direction compactified. The Matrix theory is
formulated in terms of the maximally supersymmetric $U(N)$ gauged quantum
mechanics. The momentum along the compactified
direction is identified with the integer $N$. By taking the large $N$
limit, it is hoped a complete definition of uncompactified M-theory
may be recovered. The Matrix formulation of M-theory is reviewed in \banks.

Since Matrix theory gives us a nonperturbative formulation of quantum
gravity, it should allow us to answer the long-standing questions in
quantum black hole physics. Finding a statistical interpretation for
black hole entropy is one of these important questions. Much progress
has been made in this area in recent years, with the use of D-brane
technology in string theory to give the first microscopic calculation
of the entropy for certain supersymmetric and near-supersymmetric
black holes \strovaf. However these techniques have not been helpful in the
study of black holes far from supersymmetric configurations, where it
seems that string perturbation theory breaks down. Matrix theory
should be useful in this situation. A number of recent
works \matrixbh\ have  recovered the scaling of the black hole entropy
with horizon area via Matrix theory, however an exact treatment
leading to the full Bekenstein-Hawking formula for the black hole
entropy has been lacking.

In the present work we will use the fact that, in the large $N$ limit,
Matrix theory can be thought of as an induced theory of
gravity, to map the counting of black hole microstates to the original
Gibbons-Hawking calculation \gibhawk\ of the thermodynamic black hole
entropy. In the calculation of \gibhawk, the black hole entropy arises
from a boundary term that appears in the spacetime action, which is
evaluated on a sphere of large radius surrounding the black hole. 
In this limit, the
linearized approximation is valid, and it is precisely these terms
that are induced via a one-loop calculation in Matrix theory.
Non-renormalization theorems protect these terms from receiving corrections
higher order in the coupling \sethi. To complete the calculation, we assume 
Matrix theory corresponds to the discretized light-cone quantization of a theory with 
eleven-dimensional Lorentz invariance, and apply the Noether procedure to obtain 
the non-linear form of the effective action. This allows us to obtain a relation between the
periodicity in time at infinity and the black hole mass, by demanding the absence of
conical singularities on the event horizon. As we will explain, this allows us
to map the counting of black hole microstates in Matrix theory to
the Gibbons-Hawking calculation of the Bekenstein-Hawking entropy.
This is a
realization of the idea that black hole entropy naturally has a
statistical interpretation in any theory where gravity is induced by
integrating out degrees of freedom at one loop \refs{\fiola,\jacobson}.

\newsec{Collective Fields for Large $N$ Matrix Theory}

Kabat and Taylor \kabat\ have defined a matrix analog of the stress
energy
tensor, and electric and magnetic currents of the three-form gauge
field of M-theory. Related results also appear in \chepelev.
The components of the stress energy tensor, 
which we will need in the
following, are given by the formulae
\eqn\mstress{
\eqalign{
T^{--} &= {1\over R} \str {\CF \over 96} \cr
T^{-i} &= {1\over R} \str \biggl( \half \dot \mX^i \dot \mX^j
\dot \mX^j +{1\over 4} \dot \mX^i F^{jk} F^{jk} + F^{ij} F^{jk}
\dot \mX^k \biggr) \cr
T^{+-} &= {1\over R} \str \biggl( \half \dot \mX^i \dot \mX^i +
{1\over 4} F^{ij} F^{ij} \biggr) \cr
T^{ij} &= {1\over R} \str \biggl( \dot \mX^i \dot \mX^j + F^{ik}
F^{kj} \biggr)\cr
T^{+i} &= {1\over R} \str \dot \mX^i \cr
T^{++} &= {1 \over R} \Tr \Bone~,\cr}
}
where $\str$ denotes a trace over $N\times N$ matrices, symmetrized
over products of $F^{ij}$. Here $\mX^i$ are the $N\times N$ matrices,
$F_{0i} = \p \mX_i / \p t$, $F_{ij} = i [\mX_i , \mX_j]$ and
\eqn\fiseq{
\eqalign{
\CF &= 24 F_{0i} F_{0i} F_{0j} F_{0j} + 24 F_{0i} F_{0i} F_{jk} F_{jk}
+ 96 F_{0i} F_{0j} F_{ik} F_{kj} \cr & \qquad 
+ 24 F_{ij} F_{jk} F_{kl} F_{li} - 6 F_{ij} F_{ij} F_{kl} F_{kl} ~.\cr}
}
We will actually need the full matrix
expressions, which we obtain by dropping the trace in \mstress, and
which we denote by $T_M^{\mu \nu}$. The following conventions will be
used:
$\mu, \nu = 0,\cdots ,10$ denotes eleven-dimensional spacetime indices
with signature $(-+\cdots +)$; $i,j=1,\cdots,9$ are transverse spatial
indices; $\eta_{\mu\nu}$ denotes the
eleven-dimensional Minkowski metric; 
$x^\pm = {1\over \sqrt{2}}(x^0\pm x^{10})$; $x^- \sim x^- + 2\pi
R$. The time parameter of the Matrix theory $t$ is to be identified
with $x^+$ and $\dot X= dX/dt$.

We now wish to use the expressions \mstress\ to define a collective
field
which describes low-energy excitations about a smooth background, in
a linearized approximation. 
Let us first define the quantities
\eqn\cstress{
T^{\mu \nu}(x^i,t) = \int \bigl( {{d k_\perp}\over 2\pi} \bigr)^9 e^{i
k_i x^i} \str e^{-i k_i \mX^i} T^{\mu\nu}_M(t) ~.
}
We then define our collective field $h^{\mu\nu}(x^i,t)$ via the
equations
\eqn\cmetric{
\eqalign{
\bar h^{\mu \nu} &= h^{\mu \nu} - \half \eta^{\mu\nu} h^{\rho}_\rho
\cr
 \nabla^2 \bar h^{\mu \nu}(x^i,t) &= 
-16 \pi G T^{\mu\nu} (x^i,t) ~,\cr}
}
where $\nabla^2$ only acts on the transverse coordinates. Here
$16 \pi G = (2\pi)^8 l_p^9$ defines the Planck length, and we take
$2\pi l_p^3 = R$. 

The key point that follows from \kabat\ is that, in the large $N$
limit,
$h^{\mu \nu}$ may be
identified with the spacetime metric
$g_{\mu\nu}= \eta_{\mu\nu}+h_{\mu\nu}$, at least to the extent that
the
linearized gravity equations may be trusted, $|h_{\mu\nu}| \ll 1$. 
To see this from the
matrix approach, one considers a block diagonal background for the
matrices $\mX^i$, which consists of two nontrivial blocks $\hat \mX$
and $\tilde \mX$, and integrates out the off-diagonal degrees of
freedom to give a one-loop effective action. This
yields a potential that takes the form
\eqn\mpot{
V= -{15 R^2 \over 4 r^7} \bigl( \hat T^{\mu\nu} \tilde T_{\mu\nu} -
{1\over 9} \hat T^{\mu}_\mu \tilde T^{\nu}_\nu \bigr)~,
}
where $r \equiv \sqrt{\sum (\hat x^i-\tilde x^i)^2}$, 
and $\hat T$ and $\tilde T$ are
obtained by substituting $\hat X$ and $\tilde X$ respectively into
\mstress.
This agrees with the tree-level result obtained via a classical
supergravity calculation, in the sector with zero longitudinal
momentum transfer. This is the sector relevant in the present work, as
we will consider the limit where $r \gg R$, when the $x^-$ dependence
drops out of the quantities of interest.
We may therefore regard Matrix theory as a
realization of induced linearized gravity. 

The idea of inducing gravity via integrating out a set of field
theory degrees of freedom in a nontrivial background spacetime has a
long history \sakharov. Here the difference is that we have induced
gravity in eleven dimensions by integrating out a set of quantum
mechanical degrees of freedom, rather than eleven-dimensional field
theoretic degrees of freedom. In fact we have only induced linearized
gravity -
the full
nonlinear supergravity equations of motion are only expected to be
recovered in an all orders treatment of the matrix quantum mechanics. 

\newsec{Counting Black Hole Microstates}

We may now use this set of observations to give a microscopic
derivation of the Bekenstein-Hawking entropy for a black hole.
In fact, it has already been pointed out \refs{\fiola,\jacobson}, that
any theory of induced gravity will naturally give a statistical
interpretation for black hole entropy, and we may apply that basic
observation here, although there are some important differences in
the way this comes out of Matrix theory, as will be seen in the
following. Black hole entropy in theories of induced gravity has been
further studied in \fursaev.

Let us see how the computation of the black hole entropy proceeds in this
framework. We will follow the logic of Gibbons and Hawking \gibhawk,
and continue to imaginary time, which is periodically identified with
period $\beta$. Physically, this should be thought of as a black hole
in equilibrium with a heat bath at inverse temperature $\beta$. The
heat bath will carry an entropy extensive in the size of the system.
However we may always isolate the contribution of a black hole by
taking the large mass limit, provided the size of the system is
suitably adjusted to prevent finite size effects becoming important.

The quantity $\beta$ will later be determined by a
self-consistency argument. To represent a black hole configuration in
spacetime, we impose the condition that the field $h_{\mu\nu}$
satisfy the linearized gravity equations appropriate to an object
of mass $M$ at rest. 

For simplicity, we consider the case with no angular momentum and no
electric or magnetic charges, but the generalization to non-zero
angular momentum and electric charge is straightforward, using the
matrix expressions for these quantities appearing in \kabat. Including
magnetic charge is rather more subtle as the matrix expressions for
the magnetic currents are not well understood. 

Because the $x^-$ direction is compactified, 
we must
have $M/\sqrt{2} = N_{bh}/R$ for some integer $N_{bh} <N$.
Asymptotically, the field $h_{\mu\nu}$ has non-zero components
\eqn\asyh{
\eqalign{
h_{00} &= {2 \hat M \over r^{7}} \cr
h_{ij} &= \delta_{ij} {\hat M \over {4 r^{7}}} ~,\cr}
}
where 
we have defined
\eqn\massparam{
\hat M = {5 G M \over 24 \sqrt{2} \pi^4 R}~.
}
%{16 G M \over \sqrt{2}\pi (D-1)(D-2) A_{D-4} R} \cr
%A_{D-1} &= {2 \pi^{D/2} \over \Gamma(D/2)} \cr}

The effective action that generates the linearized equations of motion
for the field $g_{\mu} = \eta_{\mu\nu}+h_{\mu\nu}$ takes
the usual Einstein-Hilbert form, plus a boundary term \gibhawk,
\eqn\gibhac{
\hat I= -{1\over 16\pi G} \int_{\CM} \sqrt{-g} \CR - {1\over
8\pi G}
\int_{\p \CM} \sqrt{-\hat  h}(\CK-\CK_0)~,
}
where $\CM$ is the eleven-dimensional space, $\CK$ is the second
fundamental form of $\p \CM$ in the metric $g$, and $\hat h$ is the
induced metric on $\p \CM$.
$\CK_0$ is the second fundamental form of the boundary
embedded in flat space. It is to be understood that these curvatures
are evaluated in the linearized gravity limit.
Although the boundary term does not affect the
equations of motion, it is necessary for a well-defined action,
which must depend on at most single derivatives of the metric.

We now wish to compute the number of matrix states that give rise to
a metric of the form \asyh, with $h_{\mu\nu}$ related to the matrix
variables by equation \cmetric. This corresponds to evaluating the
partition function
\eqn\indpart{
\Tr e^{-\beta_M H} = e^{-\hat I +\cdots}~,
}
using the low-energy effective action,
where in the trace on the left we restrict to states that give rise
to an expectation value for $h_{\mu\nu}$ as in \asyh. 
Here $H$ is the Matrix theory
Hamiltonian. The additional terms denoted $\cdots$ on the right-hand
side arise from  higher order quantum corrections to the leading term
in the effective action. We assume, at least in the large $N$ limit,
that the gravitational coupling is not renormalized beyond one loop in
the matrix quantum mechanics.\foot{Non-renormalization theorems have now been 
established that  protect these terms from corrections at higher order in the 
coupling \sethi.}
$\beta_M$ is related to the
periodicity in time of the black hole solution by
$\beta_M = \beta /\sqrt{2}$. Since we wish the states to be periodic
in $x^0$ with period $\beta$, we must project onto states invariant
under an additional shift of $\beta_M$ in the $x^-$ direction as we
shift by
$\beta_M$ in the $x^+$ direction.

The microscopic entropy may then be computed by evaluating the
right-hand side of \indpart\ and is given by
\eqn\enteq{
S= \beta \vev{E} - \hat I~,
}
where the energy is
\eqn\eneq{
\vev{E} = {\p \hat I \over \p \beta}~.
}
We have thus mapped the counting of states in Matrix theory onto the
original Gibbons-Hawking calculation of the thermodynamic black hole
entropy. It is crucial here that only the linearized
part of the supergravity theory contributes to the action $\hat I$
through the boundary term, which is evaluated on a sphere of large
radius. In the limit that the radius $r \gg R$, the $x^-$ dependence
of the integrand in the boundary term drops out. Note we must keep $R$ 
finite for this argument to work.
Precisely this contribution is induced by a one-loop
calculation in Matrix theory, and therefore corresponds to a counting
of microstates. 
It remains then to evaluate the right-hand side of the equation
\indpart. This follows
the usual Gibbons-Hawking calculation, and therefore works for an
arbitrary black hole, or black p-brane.

Finally we must fix the value of $\beta$. This should be regarded as
another boundary condition that fixes the kind of spacetime states we
wish to count. It may be seen that the form of the metric \asyh,
inevitably leads to an event horizon where the linearized
approximation to supergravity breaks down. Fortunately, the
evaluation of $\hat I$ did not depend on the details of the metric in
this region. If we assume that Matrix theory is the discretized light-cone
quantization of
a theory which at least at low energies has
eleven-dimensional Lorentz invariance\foot{It should be emphasized 
this assumption
of Lorentz invariance is not proven. However,
it has been argued \seiberg, that if a Lorentz invariant eleven-dimensional
description of M-theory does exist at the quantum level, it must be 
described in discretized light-cone formulation by Matrix theory.}, 
it follows that the nonlinear
completion of \gibhac\ must take the form of eleven-dimensional
supergravity (plus appropriate boundary terms). 
One
can make an argument analogous to \deser\ to see that once
the effects of gravitational self-interaction are included, 
the full nonlinear supergravity equations are reproduced by 
following the Noether procedure.
These nonlinear equations then determine the correct behavior of the metric
near the horizon. The inverse temperature is determined by
demanding that
the metric, which we want to identify with the
physical metric, does not contain conical singularities at the
horizon. While states with singularities at the horizon are certainly
present in Matrix theory, we should not regard such states as black
holes.

As an example,
let us consider for simplicity a black string in eleven
dimensions, which lives in the $(x^0,x^{10})$ plane. As shown in
\laflamme, the black string will only be stable if $R$ is much less
than the horizon radius.
Asymptotically the metric takes the form \asyh, which yields an action
\eqn\ghaction{
\hat I = {\pi^3 \beta R \hat M \over 27 \sqrt{2} G}~.
%{A_{8}\beta \hat M\over 8\pi G}
}

For the black string case the complete nonlinear equations of motion
give a metric with no conical singularities for
\eqn\hawkt{
\beta = {4\pi \over {7}} (2\hat M)^{{1\over {7}}}~.
}

Substituting into \enteq\ we find
the final formula for the statistical entropy of black string
microstates from Matrix theory
\eqn\entrop{
S= {A\over 4 G}~,
}
where $A$ is the area of the event horizon. This of course agrees with
the thermodynamic entropy of Bekenstein and
Hawking.

\newsec{Comments}

In the this letter we have shown the counting of black hole
microstates in Matrix theory is mapped onto the Gibbons-Hawking
calculation of the black hole entropy. We have presented the
computation for M-theory with a null direction compactified. The
calculation generalizes in a completely straightforward way to
M-theory further compactified on $T^d$ for $d\leq 5$, when Matrix
theory has a description in terms of a $d+1$-dimensional quantum field
theory with $U(N)$ gauge symmetry.

For M-theory compactified on tori with $d>5$ we run into the problem
that the $U(N)$ matrix fields are not the only degrees of freedom
present \seiberg. Nevertheless, it is natural to expect that the degrees of
freedom giving rise to black hole entropy will arise from $U(N)$
matrix fields in an analogous way. Likewise there are technical
complications when we try to compactify M-theory on curved
backgrounds \ooguri, but we expect the present picture will carry over to that
case as well.

It should be noted that the usual interpretation of black hole entropy in
induced gravity \refs{\fiola,\jacobson} is as 
entanglement entropy between quantum fields outside and
inside of the 
black hole horizon. In Matrix theory, the event horizon emerges
as a collective effect in the large $N$ quantum mechanics. There
does not appear to be a natural definition of the
horizon location directly in terms of the matrix variables. Therefore
it seems most appropriate to think of the entropy in Matrix theory as
a direct
counting of states rather than measuring quantum entanglement.

\bigskip

\centerline{\bf Acknowledgments}

I wish to thank A. Jevicki and A. Strominger for helpful discussions.
This research was supported in part by DOE grant
DE-FG0291ER40688-Task A.

\listrefs
\end